# Tunable Exciton-Hybridized Magnon Interactions in a Layered Semiconductor

Geoffrey M. Diederich[1,2*], John Cenker[2*], Yafei Ren[3*], Jordan Fonseca[2], Daniel G. Chica[4], Youn Jue Bae[4], Xiaoyang Zhu[4], Xavier Roy[4], Ting Cao[3], Di Xiao[3,2#], Xiaodong Xu[2,3#]

[1]Intelligence Community Postdoctoral Research Fellowship Program, University of Washington, Seattle, WA, USA
[2]Department of Physics, University of Washington, Seattle, WA, USA
[3]Department of Materials Science and Engineering, University of Washington, Seattle, WA, USA
[4] Department of Chemistry, Columbia University, New York, NY 10027, USA
*These authors contributed equally to this work

[#]Correspondence to: dixiao@uw.edu; xuxd@uw.edu

**Abstract:** The interaction between distinct excitations in solids is of both fundamental interest and technological importance. One example of such interactions is coupling between an exciton, a Coulomb bound electron-hole pair, and a magnon, a collective spin excitation. The recent emergence of van der Waals magnetic semiconductors[1] provides a powerful platform for exploring these exciton-magnon interactions and their fundamental properties, such as strong correlation[2], as well as their photo-spintronic and quantum transduction[3] applications. Here we demonstrate precise control of coherent exciton-magnon interactions in the layered magnetic semiconductor CrSBr. We show that by controlling the direction of applied magnetic fields relative to the crystal axes, and thus the rotational symmetry of the magnetic system[4], we can tune not only the exciton coupling to the bright magnon, but also to an optically dark mode via magnon hybridization. The exciton-magnon coupling and associated magnon dispersion curves can be further modulated by applying a uniaxial strain. At the critical strain, a dispersionless dark magnon band emerges. Our results demonstrate unprecedented control of the opto-mechanical-magnonic coupling, and a step towards the predictable and controllable implementation of hybrid quantum magnonics[5–11].

**Main text**

Magnons, the quanta of spin waves, have recently become an attractive component of hybrid quantum devices due to their ability to carry information with no resistive heating[12], to perform wave-based information processing on nanometer scale elements[13], and to allow simpler device architectures with no electrical contacts[14]. However, to implement magnonic information into a hybrid system, it is necessary to have efficient coupling between magnons and external information carriers, such as optical photons, via magnon-exciton interactions. In order to write and process information efficiently, it is highly desirable that these interactions are tunable. To that end, magnetic fields have been used extensively to tune magnons[4,15–23] and recent works have shown that strain can be an effective tool to generate and control magnons via magnetostriction[24,25]. These approaches present avenues for *in situ* tuning of the magnonic properties, ideal for the programmable interactions necessary in hybrid quantum systems.

CrSBr is a van der Waals semiconductor with an A-type layered antiferromagnetic (AFM) order[26,27], where the individual monolayer is ferromagnetic (FM) with an in-plane easy axis (along the **b** axis, Fig. 1a) and the interlayer coupling is antiferromagnetic. Previous work has shown that the exciton energy in this material depends sensitively on the interlayer magnetic configuration; namely, the spin alignment controls the interlayer hopping, which in turn controls the band gap and hence the exciton resonance energy (Fig. 1a)[28]. Specifically, the spin-dependent portion of the exciton energy is proportional to $\boldsymbol{S}_1 \cdot \boldsymbol{S}_2$, where $\boldsymbol{S}_1$ and $\boldsymbol{S}_2$ are the magnetization in adjacent layers.[28] This strong coupling between the exciton resonance and interlayer spin alignment is the underlying mechanism of the recent observation of long-lived magnons, measured by probing the exciton when CrSBr is subject to ultrafast laser excitation under a fixed out-of-plane magnetic field[3]. In addition, uniaxial strain has been shown to be a powerful knob for tuning the magnetic anisotropy of CrSBr as well as the interlayer magnetic exchange interaction, which can flip sign under appropriate strain[29]. Thus, CrSBr is an ideal model system to explore highly tunable exciton-magnon interactions.

In this letter we demonstrate control over the coherent coupling between excitons and both bright and dark magnon modes in CrSBr via external magnetic fields ($\mu_o\mathbf{H}$) and uniaxial strain ($\varepsilon$). To measure this coupling, we performed transient optical reflectivity measurements of an exfoliated thin bulk CrSBr flake. A pump pulse centered at 2.66 eV first excites the magnons with above optical bandgap photons. The probe is set near the exciton resonance of 1.33 eV. The pump and probe pulses carry 1 and 0.1 $\mu J/cm^2$ at the sample, respectively. The experimental temperature is 35 K, far below the Neel temperature of ~132 K[30] (see methods for additional experimental details).

Figure 1b shows the transient optical reflectivity versus pump-probe delay at three selected $\mu_o\mathbf{H}$ strengths applied along the **a** axis, i.e. when CrSBr is in the canted AFM state. In all main text figures, we have subtracted the electronic decay signal to emphasize oscillatory components (see Extended Data Fig. 2 for raw data, and Supplementary Text S1 and S2). The signals show strong oscillations due to the spin wave modulated exciton energy, consistent with reported exciton-magnon coupling in CrSBr with $\mu_o\mathbf{H}$ applied along the **c** axis[3]. As shown in Fig. 1a, due to the strong modulation of the exciton by the spin alignment[28], the exciton resonance energy oscillates at the spin wave precession frequency[3]. Essentially, the exciton 'senses' the precession via its spectral shift, acting as a proxy for the spin wave which is written as an amplitude modulation onto the electronic decay signal. This allows direct access to the microwave magnon excitation via optical photons without any specialized magneto-optical techniques, such as time-resolved magneto optical Kerr effect (TR-MOKE) spectroscopy. A comparison of transient reflectivity and TR-MOKE for our samples can be found in (Extended Data Fig. 1).

The oscillation frequency and exciton-magnon coupling strength are evidently dependent on the magnetic field strength. To measure this dependence, we performed time-resolved measurements as a function of $\mu_o H$ (raw data in Extended data Fig. 2). The Fourier transform of the data yields a single magnon mode corresponding to the optical magnon branch, whereas the acoustic branch remains completely dark as shown in the intensity plot of the magnon mode versus $\mu_o H$ and frequency (Fig. 1c). In addition, the brightness of the optical branch varies continuously with $\mu_o H$, reflecting the tuning of the exciton-magnon interaction strength. The data in Fig. 1c is qualitatively similar to magnon spectra recently measured in CrSBr with out-of-plane field dependence[3]. However, our field is applied in-plane, along the intermediate axis of the crystal, resulting in a magnon spectrum with distinct field dependence.

To understand the tunable magnon brightness, we recall that CrSBr has triaxial magnetic anisotropy, i.e., an easy-axis along the **b** axis, an intermediate **a** axis, and a hard **c**-axis. With $\mu_o H \parallel$ **a**, all spins cant and the resulting magnetic state is invariant under a $C_2$ rotation about the **a** axis. This symmetry dictates that the optical and acoustic branches must have opposite parity as depicted in Fig. 1d. For the acoustic (even) mode, the net spin precesses around the **a** axis and the relative angle between $S_1$ and $S_2$ is fixed with negligible effect on the exciton energy. This mode therefore cannot be directly read out by transient optical reflectivity near the exciton resonance and does not show up in the Fourier transform, i.e. it is optically dark. The optical (odd) mode is the opposite: the net spin is along the **a** axis with a time varying amplitude and the relative angle between $S_1$ and $S_2$ is time-dependent with a period the same as that of the spin wave. This produces a periodic modulation of the exciton gap and is thus optically bright due to its strong coupling to the exciton (Fig. 1e), consistent with previous reports[3]. The arrows shown above and below Fig. 1e provide the relative orientation of the $S_1$ and $S_2$ vectors for both magnons modes at various points in the magnon precession, providing direct connection between the spin wave phase, the magnetic state, and the exciton shift at various points in the oscillations. The calculated dark ($|D\rangle$) and bright ($|B\rangle$) modes are shown by the white dashed lines in Fig. 1c. The magnitude of the exciton shift in Fig. 1e was estimated by comparing the exciton overlap with the probe pulse spectrum to the modulation depth of the signal, and matches the exciton shifts measured previously[3].

Since the spin-dependent portion of the exciton energy $E \propto \cos(\theta(t)) = \cos(\theta_0 + \delta\theta(t))$, the modulation of the exciton energy due to magnons must be proportional to $dE/d\theta_0 \propto \sin(\theta_0)$. Here, $\theta(t)$ is the time-dependent angle between $S_1$ and $S_2$, $\theta_0$ is its equilibrium value and $\delta\theta(t)$ is induced by the spin wave. Tuning $\theta_0$ by $\mu_o H$ is thus a control of the exciton-magnon interaction strength. The exciton-magnon coupling vanishes in the collinear AFM state with $\theta_0 = 180°$. As $\mu_o H$ increases, the relative angle $\theta_0$ decreases. The magnon brightness increases until $\theta_0$ reaches 90°, then starts decreasing until the system reaches the FM state, where the exciton-magnon interaction again vanishes at $\theta_0 = 0°$.

It is clear from the above discussion that, constrained by two-fold rotational symmetry, the dark acoustic magnon mode cannot couple to the exciton directly with $\mu_o H \parallel$ **a** axis. The same argument also applies with $\mu_o H \parallel$ **c** axis. However, coupling can be turned on when $\mu_o H$ is tilted away from the crystal axes. The underlying mechanism is the coherent hybridization between the two magnon modes induced by the rotational-symmetry breaking magnetic field, as shown by recent microwave absorption experiments on $CrCl_3$[4]. To measure this hybridization, we performed transient optical reflectivity measurements with an applied in-plane field that deviates from the **a** axis by an angle $\theta_{ab}$ = 2 degrees (Extended Data Fig. 3). Its Fourier transform yields the magnon dispersion curve in Fig. 2a. In stark contrast to the single mode magnon spectrum at $\mu_o$**H** ∥ **a** (Fig. 1e), the magnon spectrum exhibits a pronounced avoided crossing as the magnetic field is varied, demonstrating the coherent hybridization of the bright and dark magnon modes.

Figure 2b shows the time resolved optical reflectivity at three selected magnetic fields. Intuitively, the magnon modes can be written as $\alpha|D\rangle \pm \beta|B\rangle$, where the coefficients α and β can be controlled by the magnetic field. When the two modes are off-resonance, the magnon mode is dominated by either the dark or bright modes, as observed by the single frequency damped oscillation in the top and bottom panels of Fig. 2b. At the avoided crossing ($\mu_o$**H** ~0.53 T, middle panel of Fig. 2b), the bright and dark magnon modes are coherently hybridized with nearly equal contributions (i.e. $\alpha = \beta$), leading to a beating pattern with near unity modulation depth. These coherently hybridized modes can, alternatively, be thought of as the new normal modes of the system under a symmetry breaking field. An interesting feature in the data is the apparent increase in the magnon lifetime at the avoided crossing, where the linewidth decreases by a factor of 3 with respect to that of the unhybridized optical magnon. Exploitation of this property could prove useful for spintronics applications requiring long lifetimes and transport lengths. The corresponding spin wave motion is depicted in Fig. 2c in which one of the spins is fixed while the other spin precesses. In the hybridized modes, the relative angle between the spins is time dependent and produces a modulation of the exciton energy analogous to that depicted in Fig. 2d.

The coupling strength between the bright and dark magnon modes, manifest in the frequency splitting, Δ, at the avoid crossing, can be controlled by $\theta_{ab}$. Figure 3a shows the magnon dispersion curves at selected $\theta_{ab}$, obtained by Fourier transformation of the transient optical reflectivity (Extended Data Fig. 4). The increased splitting shows that increasing $\theta_{ab}$ leads to stronger coupling of the bright and dark modes. The increased coupling strength then results in a brighter dark magnon, evidenced by the tails that extend from the avoided crossing and follow the dispersion of the dark mode. Our simulations in Fig. 3b excellently reproduce both features as well as the relative brightness of the magnons (see the methods section for details). The dependence of the magnons over the full range of $\theta_{ab}$ can be found in (Extended Data Fig. 5).

Figure 3c plots the extracted hybridization strength, Δ, as a function of $\theta_{ab}$, showing a linear dependence that is also reflected in the increasing beat frequency of the time domain traces at the avoided crossing (Fig. 3d). The splitting can be as large as half of the low energy magnon mode, demonstrating the strongly tunable coupling. In addition, we extracted the dark magnon mode amplitude at fields far from the avoided crossing ($\mu_o$**H** = 0.3 T), which represents the coupling strength between the dark magnon and the exciton (Fig. 3e). It is clear that the application of a symmetry-breaking magnetic field enables on/off switching of the exciton coupling to the dark mode, tunability in the bright-dark magnon hybridization, and tuning of the relative strengths of the two magnon modes in the measured signal. The properties we have found extend the functionality of the exciton magnon coupling[3] in CrSBr and enable device implementation. Such control of the magnon coupling allows for the controlled generation of a variety of two magnon states, useful for the state initialization required for information processing.

We extend our control scheme by applying a uniaxial strain to realize *in situ* control of the exciton-magnon interactions via mechanical tuning of the magnetic exchange interactions and anisotropy. To achieve this, we clamped the CrSBr flake to a flexible polyimide substrate and applied tensile strain using a piezoelectric strain cell (Fig. 4a). The efficient clamping scheme enables the application of strains up to ~ 2%, as determined by Raman spectroscopy (Extended Data Fig. 6 and Supplementary Text S3). Figure 4b shows the magnon spectra at selected strains applied along the crystal **a** axis with a field angle $\theta_{ab} = 9°$. Immediately obvious is the decrease in saturating magnetic field, consistent with the strain induced reduction of the magnetic anisotropy[29]. At higher strains, where the sample becomes ferromagnetic, no magnon is observed in the transient optical reflectivity spectra (Extended Data Fig. 7), consistent with the strain induced AFM to FM phase transition known to occur in CrSBr[29]. It is noteworthy that at a fixed magnetic field, e.g. $\mu_o H = 0.2$ T, increasing the strain gradually increases the coupling between the dark magnon and the exciton, further demonstrating the high degree of tunability available in this system.

A prominent effect of the strain is the strong modification of the magnetic field dependence of the magnon frequencies. As shown in Fig. 4b, the strain acts to bend the magnon dispersion curves, shifting their curvatures from having opposite signs at low strains to similar signs at high strains. Figure 4c shows simulations of the magnon dispersion versus strain, which confirms both the decrease in the saturating magnetic field as well as the change in the magnon band curvature. The fits of the magnon data enable us to extract relevant magnetic parameters; an otherwise difficult task for thin van der Waals crystals. They reveal that while the interlayer exchange, $J_2$, is significantly reduced by strain (Fig. 4d), the magnetic anisotropy difference along the **a** and **b** axes remains constant within the error of our fitting (Fig. 4e). These results show that strain applied along the **a** axis does not change the easy axis direction (i.e. it is still along **b**), and imply that the strain-induced magnetic phase transition is mainly driven via tuning of the interlayer exchange interaction.

Finally, our simulations reveal that the sign switching of the magnon curvature happens completely in the dark magnon band; causing it to pass through a constant-energy state at an applied strain of about 1.2% (white dashed lines in Fig. 4c). Signatures of field-independent magnon have recently been reported in Raman experiments on MnBi$_2$Te$_4$, arising from scattering between two magnons with equal but opposite Zeeman shifts[31]. In our single dark magnon case, the appearance of the flat magnon band with respect to magnetic field is caused by the relative orientations of the spin precession and magnetic field directions. At the critical strain, the angular momentum of the magnons is perpendicular to $\mu_o H$ and thus experiences no Zeeman shift (see supplementary Text S4). This unique magnon state with the *in situ* tunability warrants further exploration.

Hybrid magnonic systems hold vast potential for future information technologies, but that potential requires controlling the frequency behavior and coupling of magnons. Here, we have shown that these criteria are met in the 2D magnetic semiconductor CrSBr. CrSBr has been previously shown to possess strong exciton-magnon coupling[3], and we have now presented a system whose coupling between electronic, spin, and lattice degrees of freedom is both highly tunable and, considering the agreement of our data with theoretical simulations, predictable. CrSBr presents strong tunable coupling of magnons to external information carriers, providing a connection between microwave excitations and optical photons; evidence for the potential of the system for the transduction of quantum information between disparate quantum nodes[32–34]. These attributes position CrSBr as an ideal platform for encoding information and our results lay the groundwork for pushing forward hybrid magnonic systems based on 2D magnets, with CrSBr acting as a key element.

**Methods**

**Sample fabrication and strain application:** To prepare the strain substrate, transparent 20 μm thick polyimide was cut into strips. These strips were then adhered onto 2D flexure sample plates produced by Razorbill instruments using Stycast 2850 FT epoxy. The effective gap size (distance between the epoxy on either side of the gap) was less than 200 μm to enable large strains.

Bulk CrSBr crystals were grown by the same method detailed previously[35]. These crystals were exfoliated onto PDMS and thin (~ 20 nm/~25 layers) flakes were identified by optical contrast. Though thin bulk flakes were used, the magnon has weak thickness dependence[3] and our results should hold for all thicknesses greater than monolayer. The flakes were then transferred onto the epoxied polyimide strip via a dry transfer technique using a stamp consisting of a polycarbonyl (PC) film placed on top of a polydimethylsiloxane (PDMS) cylinder. The long axis of the CrSBr flake was aligned with the strain axis for consistency with the previous study. After washing off the PC film, the window clamping pattern was fabricated using standard lithography practices with a metal thickness of 7 and 70 nm Cr and Au, respectively. After clamping the CrSBr flake to the polyimide, the sample was screwed into the same symmetric three-piezo strain cell used previously[36,37] which was used for strain experiments on bulk crystals and our previous experiments on suspended CrSBr flakes. For the measurements presented in Figs. 1-3, the CrSBr was exfoliated onto silicon wafers with a 90 nm silicon dioxide layer.

**Optical measurements:** Transient optical reflectivity measurements were performed by tuning the output of a titanium sapphire oscillator to 1.33 eV, overlapped with one of the CrSBr exciton states. The output was frequency doubled and the second harmonic and fundamental were separated into pump and probe arms of the experiment by a dichroic mirror. The probe beam was sent to a retroreflector mounted on a motorized translation stage in order to produce the pump-probe delay. Each beam was sent through a waveplate and polarizer to simultaneously attenuate the beams and align their polarization to the crystal axes. The beams were recombined and sent through a 0.6 NA microscope objective onto the sample. The back-reflected beam was measured on a photodiode with a lock-in amplifier demodulating the signal at the frequency of a mechanical chopper placed in the pump arm of the experiment. To produce the time domain data, the delay stage was continuously swept at low speed while streaming data from the lock-in amplifier to the host computer at a high sampling rate (> 100 KHz), which produced time traces with 4 picosecond resolution in the data presented here. Multiple traces (4 < N < 25) were recorded and averaged, depending on the desired signal-to-noise ratio. The samples were kept at 35 K in an optical cryostat with an integrated vector magnet capable of applying fields up to 1 T along any arbitrary direction on the unit sphere. In order to determine the orientation of our applied field with respect to the crystal axes, a separate measurement was taken with the field amplitude set to the center of the avoided crossing while we varied only the field angle. Zero angle was then set to be the point where the splitting, $\Delta$, went to zero. Field angle were then set by appropriately varying the two orthogonal in-plane magnetic fields in our vector magnet.

For Raman spectroscopy measurements, 300 μW of light from a HeNe laser at 632.8 nm was focused with a 0.6 NA objective to a spot with ~ 1 μm radius. The collected light was dispersed by an 1800 groove/mm grating and detected with a liquid nitrogen cooled charge-coupled device (CCD) with an integration time of 180 seconds. BragGrate notch filters were used to filter out

Rayleigh scattering down to ~5 cm$^{-1}$. A roughly linear background originating from weak polyimide photoluminescence was subtracted to increase the accuracy of the fitting results.

**Calculation of magnon modes**: We model the magnetic properties of CrSBr by using the macrospin approximation, where the magnetization direction is assumed to be uniform within each layer. The interlayer coupling $J_2 \boldsymbol{S}_1 \cdot \boldsymbol{S}_2$ is antiferromagnetic, where $J_2 > 0$ and $\boldsymbol{S}_1$ and $\boldsymbol{S}_2$ denote the magnetizations of neighboring layers. The magnitudes of $\boldsymbol{S}_1$ and $\boldsymbol{S}_2$ are set as $S = 3/2$. The total Hamiltonian is given by

$$H = J_2 \boldsymbol{S}_1 \cdot \boldsymbol{S}_2 - g\mu_B \boldsymbol{H} \cdot \sum_{i=1,2} \boldsymbol{S}_i + \sum_{i=1,2} K_a (\boldsymbol{S}_i \cdot \hat{a})^2 - K_b (\boldsymbol{S}_i \cdot \hat{b})^2 + K_c (\boldsymbol{S}_i \cdot \hat{c})^2$$

where $\hat{b}$, $\hat{a}$ and $\hat{c}$ are the easy, intermediate, and hard axes, respectively. Since the shift of all three anisotropy energies by a constant value does not change the spin dynamics, we set $K_a = 0$, $K_{b,c} > 0$. The external magnetic field $\boldsymbol{H}$ couples to the local spins via the Zeeman coupling $-g\mu_B \boldsymbol{H} \cdot \sum_{i=1,2} \boldsymbol{S}_i$, where the Landé factor $g = 2$ and $\mu_B$ is the Bohr magneton. When the magnetic field is tilted away from the easy axis, $\boldsymbol{S}_1$ and $\boldsymbol{S}_2$ are no longer colinear and spin canting appears. The canting angles can be obtained by minimizing the total energy with respect to the orientations of $\boldsymbol{S}_1$ and $\boldsymbol{S}_2$.

To obtain the spin wave mode and their frequencies, we solve the standard Landau-Lifshitz-Gilbert (LLG) equation

$$\dot{\boldsymbol{S}}_i = -\boldsymbol{S}_i \times \boldsymbol{H}_i$$

where the effective magnetic field for the $i$-th spin is $\boldsymbol{H}_i = \nabla_{S_i} H$,

$$\boldsymbol{H}_1 = J_2 \boldsymbol{S}_2 + g\mu_B \boldsymbol{H} - 2K_b \boldsymbol{S}_1 \cdot \hat{b} + 2K_c \boldsymbol{S}_1 \cdot \hat{c}$$
$$\boldsymbol{H}_2 = J_2 \boldsymbol{S}_1 + g\mu_B \boldsymbol{H} - 2K_b \boldsymbol{S}_2 \cdot \hat{b} + 2K_c \boldsymbol{S}_2 \cdot \hat{c}$$

Calculation details are given in Supplementary Text S4.

**Acknowledgements:** This work was mainly supported by the Department of Energy, Basic Energy Sciences, Materials Sciences and Engineering Division (DE-SC0012509). Sample fabrication and optical measurements are partially supported by AFOSR FA9550-19-1-0390. Synthesis of the CrSBr crystals is supported as part of Programmable Quantum Materials, an Energy Frontier Research Center funded by the U.S. Department of Energy (DOE), Office of Science, Basic Energy Sciences (BES), under award DE-SC0019443. DGC was supported by the NSF MRSEC on Precision-Assembled Quantum Materials (DMR-2011738). This research was supported by an appointment to the Intelligence Community Postdoctoral Research Fellowship Program at University of Washington, administered by Oak Ridge Institute for Science and Education through an interagency agreement between the U.S. Department of Energy and the Office of the Director of National Intelligence.

**Author contributions:** XX conceived the project. GD performed the measurements with help from JC, JF, and YB. JC designed the strain technique and fabricated the samples. GD, JC, YR, XYZ, TC, DX, and XX analyzed the data and interpreted the results. YR, TC, and DX built the model and performed the simulations. DGC and XR grew the CrSBr crystals. GD, JC, YR, DX, and XX wrote the manuscript with input from all authors. All authors discussed the results.

**Competing Interests:** The authors declare no competing financial interests.

**Data Availability:** The datasets generated during and/or analyzed during this study are available from the corresponding author upon reasonable request.

**Figures:**

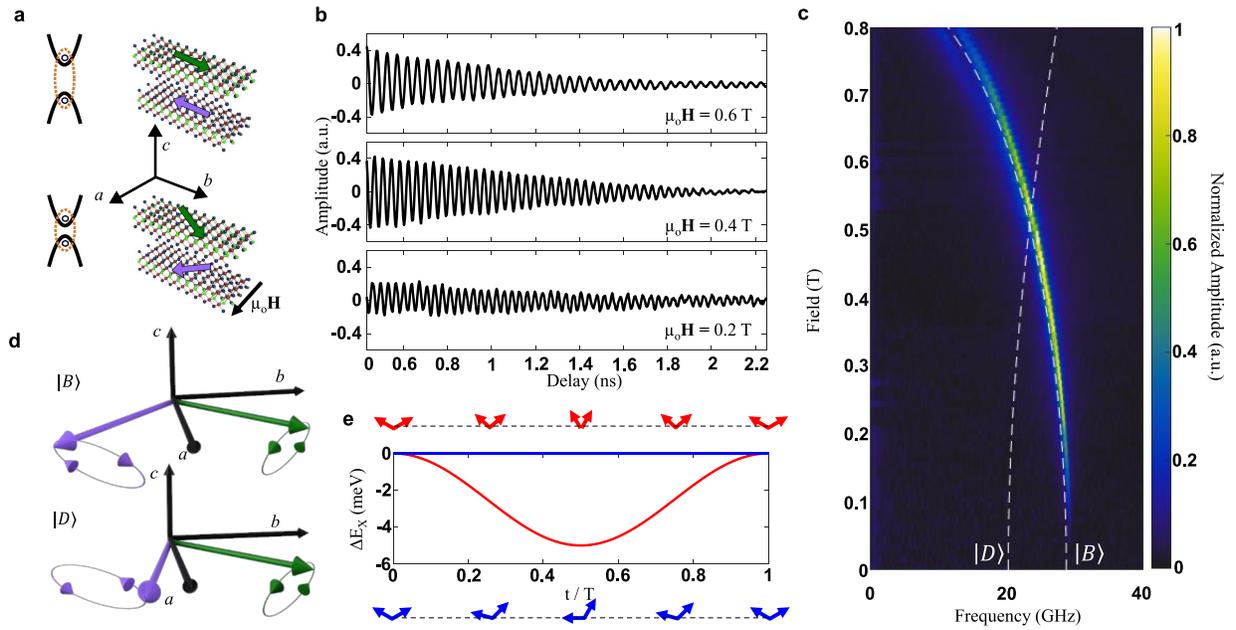

**Figure 1 | Magnetic field tuning of exciton-magnon coupling. a,** Cartoon showing the relationship between the relative spin alignment and the exciton resonance energy, where purple and green arrows represent the spins in adjacent layers. The magnetic field, $\mu_oH$, is along the crystal **a** axis in the bottom panel, leading to spin canting. **b,** Transient optical reflectivity as a function of pump-probe delay at selected magnetic field values, $\mu_oH \parallel$ **a** axis. **c,** Magnon dispersion of CrSBr, obtained by Fourier transform of the data in (**b**) and Extended Data Fig. 2. White dashed lines are the computed magnon frequencies for both the bright ($|B\rangle$) and dark ($|D\rangle$) branches. **d,** Schematics of the bright ($|B\rangle$) and dark ($|D\rangle$) magnon modes. In the bright optical magnon mode the out-of-phase precession creates a time dependent angle between the macro-spins, while the in-phase precession of the dark acoustic mode leads to a constant angle. **e,** Calculated exciton resonance shift during the precession of the optical (red) and acoustic (blue) magnon modes. The arrows denote the relative alignment of the spins throughout their precession.

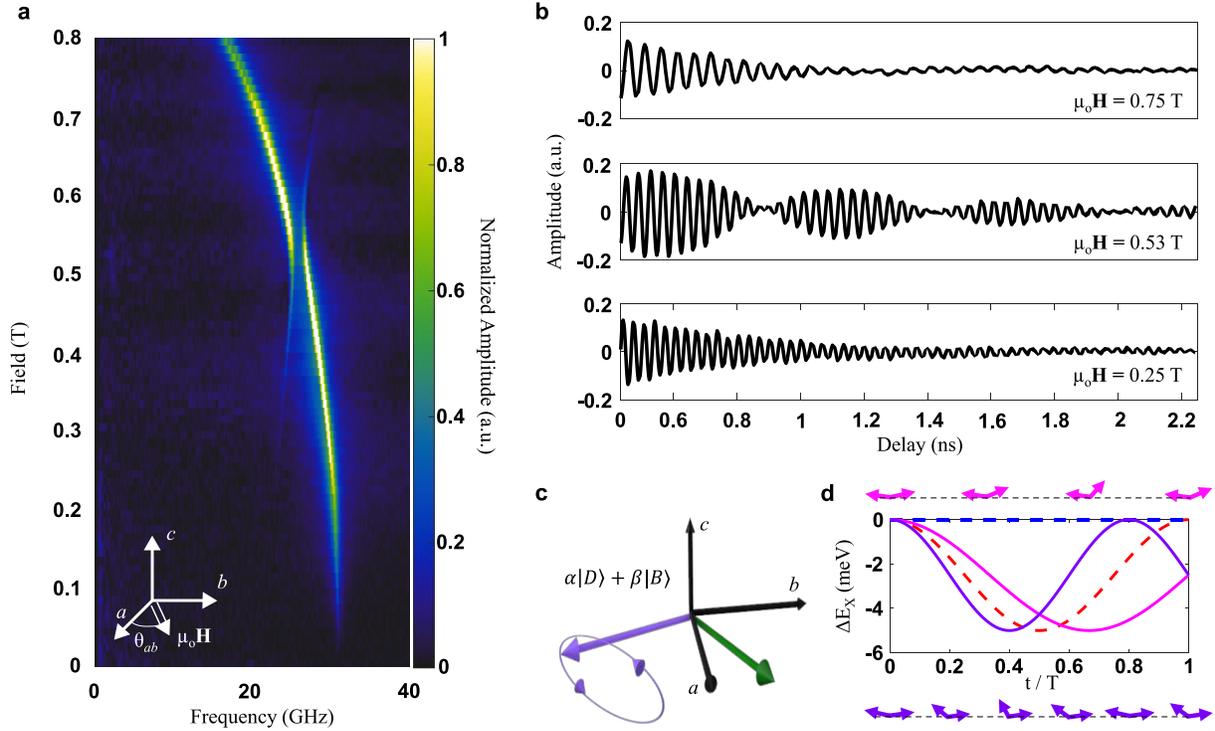

**Figure 2 | Coherent coupling between the bright and dark magnon modes. a,** Magnon dispersion with a magnetic field applied at $\theta_{ab} = 2°$ to the **a** axis. Inset shows the definition of $\theta_{ab}$. The observed avoided crossing demonstrates the coherent hybridization between the bright and dark magnon branches. **b,** Transient optical reflectivity at selected magnetic field values. The beating at 0.53T is attributed to the magnon hybridization. **c,** Spin precession for one of the fully hybridized magnon modes. **d,** Calculated exciton resonance shift under the precession of the hybridized magnon modes, shown in magenta and purple. The dashed lines correspond to the uncoupled magnon modes which are the same as Fig. 1e. The arrows show the time dependent angle between spins.

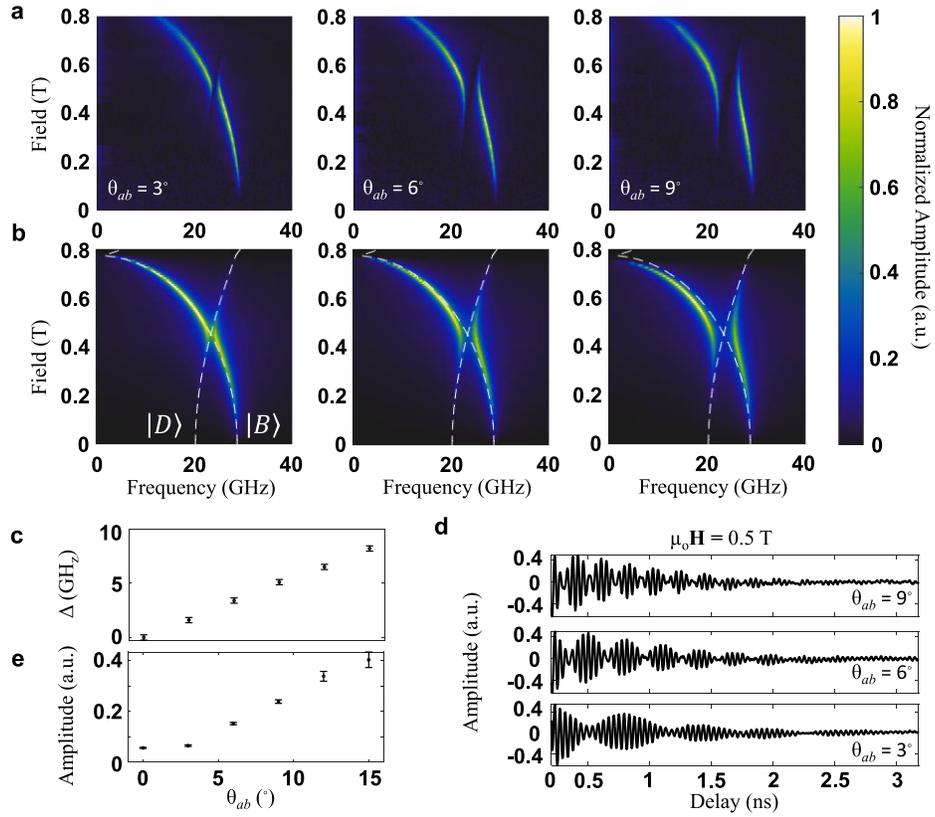

**Figure 3 | Control of exciton-hybridized magnon coupling via symmetry breaking magnetic field. a,** Magnon spectra measured at three selected magnetic field angles, $\theta_{ab}$. **b,** Simulation of the magnon dispersion and brightness for the field angle values presented in (**a**). White dashed lines are the computed magnon frequencies for both the bright ($|B\rangle$) and dark ($|D\rangle$) branches. **c,** Energy splitting of the modes at the $\mu_o H = 0.55$ T with respect to field angle. **d,** Time domain traces at the avoided crossing for each $\theta_{ab}$ presented in (**a**). **e,** Amplitude of the dark magnon mode measured at $\mu_o H = 0.3$ T with respect to field angle. Error bars in (**c**) and (**e**) represent the standard deviation of four separate measurements.

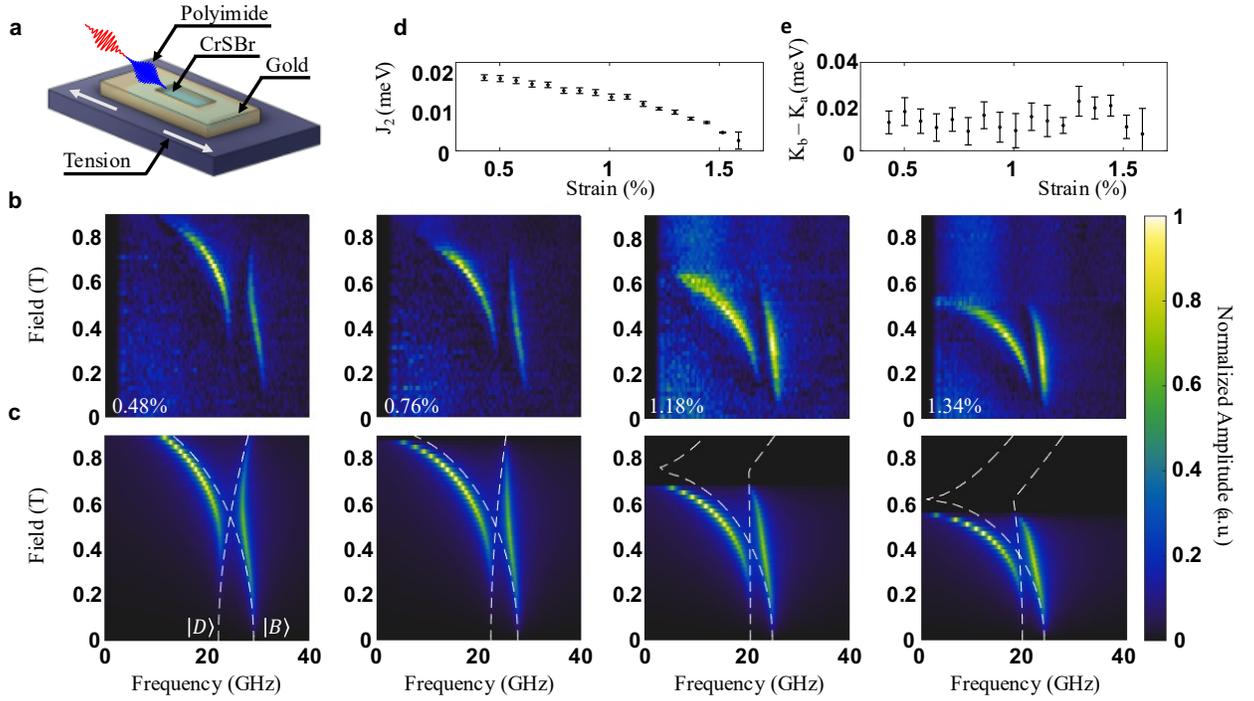

**Figure 4 | Strain tuning the exciton-magnon coupling. a,** Schematic of the strain application apparatus. **b,** Magnon spectra in CrSBr measured at selected strains applied along the **a** axis and with a field angle of $\theta_{ab} = 9°$, and **c,** corresponding magnon simulation results. White dashed lines are the computed magnon frequencies for both the bright ($|B\rangle$) and dark ($|D\rangle$) branches. **d,** Interlayer exchange, and **e,** difference in the magnetic anisotropy energies of the **b** and **a** axes as a function of strain, extracted from the simulations.

14